\documentstyle[twocolumn,psfig]{article}

\begin{document}

\begin{titlepage}
 
\begin{center}
\vspace{3.5in}
{\bf Polyakov Loops and Magnetic Screening from Monopoles in
$SU(2)$ Lattice Gauge Theory\\}
\vspace*{.5in}
John D. Stack \\
\vspace*{.2in}
{\it Theoretical Physics \\
University of Oxford \\
1 Keble Road\\
Oxford OX1 3NP \\
United Kingdom\\
and\\
 Department of Physics$^{*}$ \\
University of Illinois at Urbana-Champaign \\
1110 W. Green Street \\
Urbana, IL 61801, U.S.A. \\}
\vspace*{.4in}
{\it (presented at Lattice 96)\\}
\end{center}
\vspace*{.4in}
We present results from magnetic monopoles in $SU(2)$ lattice gauge
theory at finite temperature.  The lattices are $16^{3}\times N_{t}$,
for $N_{t}=4,6,8,12$, at $\beta=2.5115$. Quantities discussed are:
the spacial string tension, Polyakov loops, and the screening of
timelike and spacelike magnetic currents.

\vspace*{1.4in}

$^{*}$permanent address

\end{titlepage}
\vfill\eject

\pagestyle{empty}

\newpage
\pagestyle{plain}

\section{Introduction}

Magnetic monopoles found after gauge-fixing into the maximum Abelian gauge
have been successful in explaining the fundamental string tension at
$T=0$ in $SU(2)$ lattice gauge theory \cite{jssnrw}.  However, there has
been an apparent serious problem with the spacial string tension at finite
temperature  \cite{js_lat94}.
This problem has recently been resolved, and the monopole results are now
in good agreement with the full $SU(2)$ answers.  This is discussed in more
detail in a recent paper \cite{jssnrw_msum}, so here I will only mention
that the difficulty was with the method of calculation rather than monopoles
themselves.  A contribution from Dirac sheets, derived first by Smit
and van der Sijs \cite{smit}, had been omitted.  Evidently this is justified
at $T=0$, but not for temperatures above
the deconfining temperature, which corresponds to $N_{t}=8$ for the present
calculations at $\beta=2.5115$ \cite{fingberg}. 

In the remainder of this report, I will discuss the contributions of monopoles
to correlators of Polyakov loops, and the closely related question of 
how the magnetic current screens itself.

\section{Polyakov Loops}

We denote by $C(R)$ the correlation function of Polyakov loops,
$<P(R),P(0)>$, where the arguements refer to the spacial locations
of the two Polyakov loops.  For temperature $T < T_{c}$, or $N_{t} > 8$,
$C(R)$ can be used to measure a T-dependent  physical string tension, 
which is expected to approach 0 smoothly for $SU(2)$ lattice gauge theory
as $T \rightarrow T_{c}$.  
For $T > T_{c}$, or $N_{t} < 8$, we are in the 
deconfined phase and $C(R)$ determines electric screening masses.

The above remarks are for $C(R)$ in full $SU(2)$.  Here we are concerned with
how much of this physics can be obtained from monopoles.  The monopole 
contribution to $C(R)$
is readily calculated using the same method as for Wilson loops.
The naive method \cite{jssnrw} can be used, since numerically the
Dirac sheet contributions are negligible for Polyakov loop correlators.
In the following we denote the correlator of Polyakov loops calculated from
monopoles by $C_{mon}(R)$.

In Fig.~(1) we show $- \ln(C_{mon})/N_{t}$ vs. $R$ 
for the four temperatures
$T/T_{c}$ ~= 0.66, 1.00, 1.33, and 2.00 respectively, corresponding to $N_{t}$ ~=12, 8, 6, and 4.
The data points were generated by averaging over 500 widely spaced configurations
at each $N_{t}$.  The algorithm used and details of the gathering of configurations are given in
\cite{jssnrw}.
\begin{figure}
\psfig{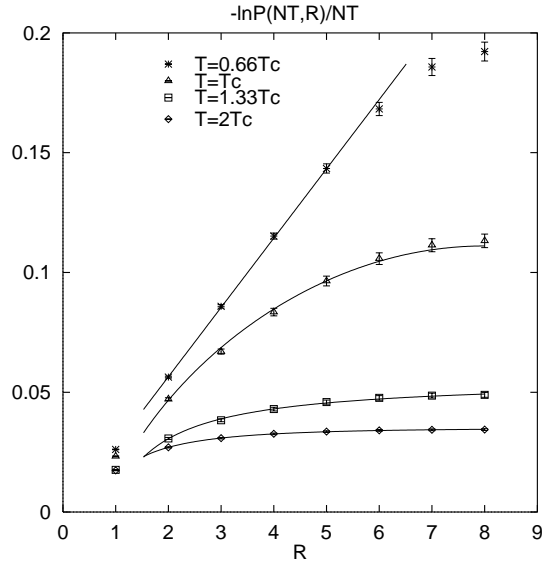}
\caption{Log of the monopole Polyakov loop correlator vs. R}
\end{figure}

First taking the case of $N_{t}=12$, or $T/T_{c}=0.66$,  the data points
are well fit by a simple linear function.  The slope of the straight line
determines the physical string tension, $\sigma_{p}(T)$. The fit gives
$\sigma_{p}(2/3T_{c})= 0.029(2)$.  We may translate this into a zero
temperature value using the formula 
$$ \sigma_{p}(0)=\sigma_{p}(T)+\frac{\pi}{3N_{t}^{2}}$$
which gives $\sigma_{p}(0)=0.036(2)$, within errors of high precision
numbers for the zero temperature string tension at $\beta=2.5115$.
So for finite temperatures $T<T_{c}$,  monopoles appear to explain the physical string tension
just as at zero temperature.  We have no other runs at finite temperature below $T_{c}$, 
but the data for $T=T_{c}$ shows curvature at all values of $R$ implying that
the physical string tension has vanished.

Turning to $T/T_{c}=1.33,2.00$, or $N_{t}=6,4$, Fig.(1) shows a very
different behavior for $-ln(C_{mon})/N_{t}$.  The expected behavior for $-ln(C)/N_{t}$ 
in full $SU(2)$ is of the general form, $A-B/R^{\gamma}\exp(-\mu(T) R)$, where
$\mu(T)$ is an electric screening mass.  At very high temperature, in the perturbative
regime, $\gamma=2$, and $\mu(T)$ can be calculated analytically, starting at
one loop.  At lower temperatures, in the non-pertubative regime, $\gamma=1$, and 
and $\mu(T)$ is determined by the lightest glueball mass in three-dimensional 
$SU(2)$ gauge theory. Our data is not extensive enough to determine $\gamma$ and
$\mu(T)$ independently in fits to $-ln(C_{mon})/N_{t}$.  Instead we explored fits
where a value of $\gamma$ was chosen, and then a minimum $\chi^{2}$ was 
searched for by varying $\mu(T)$.  For the choice $\gamma=2$, we obtain $\mu(T)=
0.0 \pm 0.1$, whereas for $\gamma=1$, the range of allowed screening masses
is slightly larger, $\mu(T) = 0.0 \pm 0.2$.  In short, although  qualitatively
$-ln(C_{mon})/N_{t}$ is of the expected general shape, quantitatively we have negligible evidence
for an electric screening mass arising from monopoles at temperatures $T > T_{c}$.
At ultra-high temperatures, this is not surprising, since the
electric screening mass is calculable from gluons in perturbation theory, and
monopoles although present, are not needed to explain it.
At lower temperatures still satisfying $T > T_{c}$, the electric screening
mass can no longer be calculated perturbatively. Nevertheless, our results suggest
that also here, its origin is not to be explained by monopoles.  

It is important to emphasize
that Polyakov correlators get contributions only from
magnetic currents in purely spacial directions.  Meanwhile a spacial Wilson loop, which 
determines the spacial string tension, receives contributions from magnetic currents
in both space and time directions. (It is the directions dual to the plane of the
generalized Wilson loop being computed that count.) 
In the next section, we will show that the time and space components of the magnetic current
behave very differently.

\section{Magnetic Screening}

We begin by reviewing the reasons for believing the magnetic current screens itself.  One
comes from the classic work of Polyakov for d=3 $U(1)$ theory with monopoles \cite{polyakov}.  This is 
relevant here, since at high temperature a $d=4$ theory effectively looks
three dimensional.
Polyakov's analysis ties an area law for Wilson loops to plasma-like 
behavior for the monopole gas, with strong Debye screening.  Another arguement for
screening comes from the Abelian Higgs model when  it is equivalent to
a type II superconductor \cite{wyld}.  Taking the ``dual" , the tube of
electric flux connecting a pair of opposite sign external charges is accompanied by strong
screening of the (magnetic) supercurrent in directions perpendicular to the 
flux tube. In both cases, confinement is accompanied by strong screening
of the magnetic charges (or currents) which cause confinement.

Returning to our finite temperature calculations, for a Wilson loop in xy,xz, or yz planes, the
magnetic current which contributes is in
tz, ty, or tx planes.  
For such purely spacial loops we do see an area law (spacial string tension).
The only component of current common in all these cases is the
time component.  This suggests strong screening of the time component of the magnetic 
current in spacial directions. 

For a Polyakov correlator, the generalized Wilson loop is in
tz, tx, or ty planes and the corresponding magnetic current which contributes is in
xy,yz, or xz planes.  
For Polyakov loop correlators we do not have
an area law. Further, the time component of the magnetic current
never contributes.  We may then have a consistent situation if strong
screening of the time component of magnetic current  is accompanied by weak, partial
screening of the spacial components.

To test these ideas, we study screening 
using a formalism explained in more detail
in \cite{jsrw}.  An infinitesimal external ``static" monopole-anti-monopole pair
is inserted
into the system of magnetic current, then the potential between this 
pair in momentum space is calculated . The magnitude of the external charges is $\kappa g$,
where $\kappa <<1$, and $g = 2\pi/e$ is the unit of magnetic charge.  For the screening of
the time-component of the current, we obtain for the screened potential $V_{\kappa}(k)$,
$$
V_{\kappa}({ k}) = 
-g^2V({ k})(1-g^2m_{44}({ k})V({ k})),
$$
where         $V({ k})$ is the $d=3$ Fourier Transform (FT) of 
         the lattice Coulomb potential, and $m_{44}({ k})$    is the
         FT of the static magnetic charge
         correlation function. 
 The  effect  of  magnetic  vacuum  polarization  on $V_{\kappa}({ k})$
 is   contained   in   
$f({ k})=m_{44}({ k})V({ k})$.

\begin{figure}
\psfig{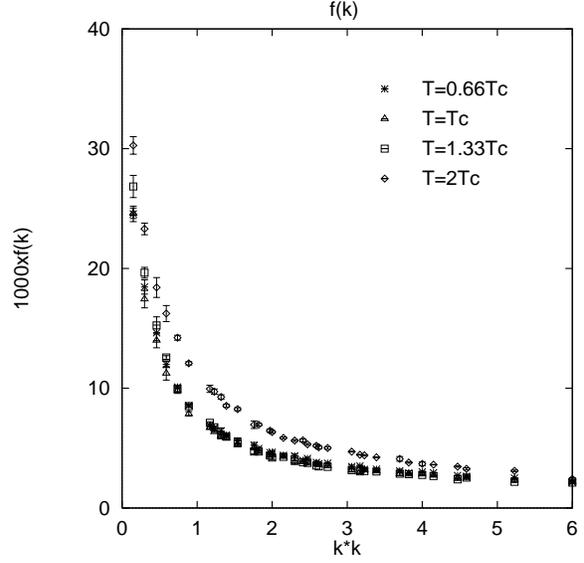}
\caption{Timelike current screening in momentum space}
\end{figure}

In Fig.(2), we show $f(k)$ vs $k^2$ for all four temperatures.
The screening is stronger ($f(k)$ larger) for $T=2T_{c}$, and rather similar for
the other temperatures, consistent with the much larger spacial string tension
for $T=2T_{c}$ \cite{jssnrw_msum}.

More quantitatively, if $V_{\kappa}$ is exponentially screened in 
position space, then
$1/f(k)$ should be  linear in $k^{2}$ for small $k^{2}$.  In Fig.(3) we 
show $1/f(k)$ vs. $k^{2}$ for the time-like current. 
\begin{figure}
\psfig{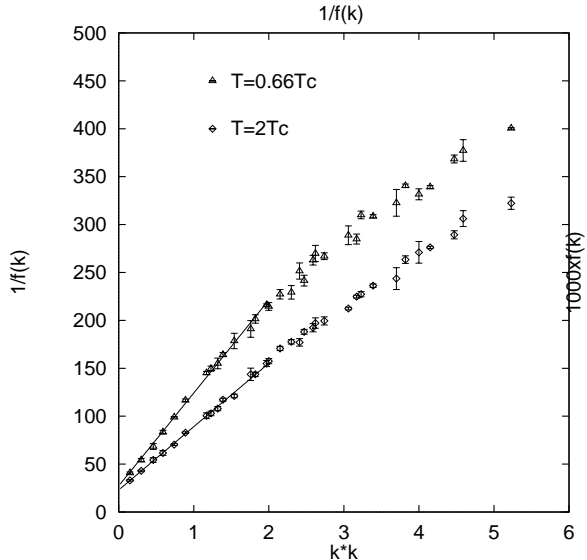}
\caption{$1/f(k)$ vs. $k^2$ for the time-like magnetic current}
\end{figure}
The slope defines the square of a
magnetic screening mass $M_{mag}^{sc}$.  Fits to the data give
$M_{mag}^{sc}=0.6(1)/a$ for $T/T_{c}=2$, and $0.5(1)/a$ for 
$T/Tc=1.33,\ 1.0,\ 0.66$
An intriguing question for the future is the relation between this
screening mass found from the time-like magnetic current and the
magnetic mass found from the gluon propogator \cite{karsch}.

If $V_{\kappa}$ has no long range part, $1/f(0)$ must equal $g^2$.
Using straight line fits to $1/f(k)$, the fitted value of
$1/f(0)=24.3 \pm 1.2$ This compares well with 
$$g^{2}=\frac{4 \pi^{2}}{e^2}=\beta \pi^{2} =24.789$$

Turning now to the screening of the spacial current, we change our viewpoint 
and regard a spacial direction as the ``static" one.  The resulting screened potential
$V_{\kappa}$ is given by a formula identical to the one above except now
the screening function 
$f({ k})=m_{xx}({ k})V({ k})$, if the static direction was chosen as the x-direction,etc for y,z.
\begin{figure}
\psfig{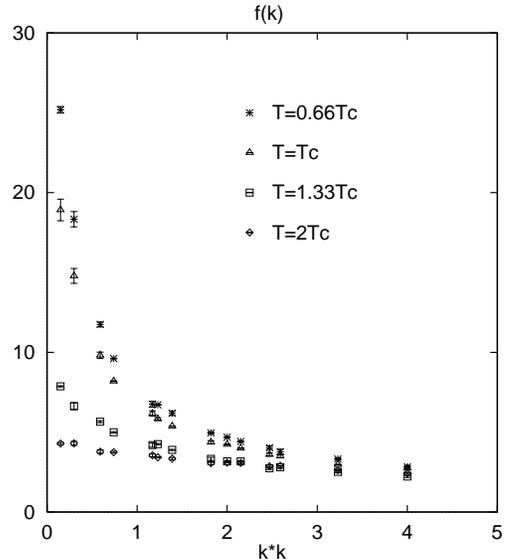}
\caption{Spacelike current screening in momentum space}
\end{figure}
In Fig.(4), we show the results averaged over the three choices x,y,z.
Comparing Figs.(2) and (4), we see that the screening of the spacial current,
in contrast to the timelike, 
{\it decreases} with temperature.  To take the most extreme case, for $T=2T_{c}$,
for the time component of the magnetic current, the screening is strongest ($f(k)$
largest), while for the spacial component of the current, the screening is
 weakest ($f(k)$ smallest).  

To summarize, we have a consistent picture of monopole
dynamics emerging from study of spacial Wilson loops, Polyakov loop correlators,
and magnetic current screening.  The spacial string tension reflects the
strong Debye-like screening of the time component of the magnetic current.  
The
Coulombic behavior (zero screening mass) of Polyakov loop correlators reflects the weak screening
of the spacial magnetic current.

This work was supported by the National Science Foundation.  
The calculations were carried out on the Cray
C90 system at the San Diego Supercomputer Center (SDSC),
supported in part by the National Science
Foundation.

\end{document}